\documentclass{ws-procs975x65}

\def\beq{\begin{equation}}
\def\eeq{\end{equation}}

\begin{document}

\title{Field theory for gravity at all scales}

\author{Michele Levi}

\address{Institut de Physique Th\'eorique, CEA \& CNRS, Universit\'e Paris-Saclay\\
91191 Gif-sur-Yvette, France\\
michele.levi@ipht.fr}

\begin{abstract}
We review here the main advances made by using effective field theories (EFTs) in classical 
gravity, with notable focus on those unique to the EFTs of post-Newtonian (PN) gravity. We then 
proceed to overview the various prospects of using field theory to study the real-world 
gravitational wave (GW) data, as well as to ameliorate our fundamental understanding of gravity 
at all scales, by going from the EFTs of PN gravity to modern advances in scattering 
amplitudes, including computational techniques and intriguing duality relations between gauge 
and gravity theories. 
\end{abstract} 


\bodymatter


\section{Introduction}

In recent years there has been a growing recognition in the all-encompassing relevance of the 
field theoretic framework beyond the quantum realm of particle physics, in which much of it has 
been conceived. Most notably, for long, in the absence of a viable quantum theory of gravity, 
the overlap between field theory and classical gravity has been largely deemed nonexistent. In 
particular, the conceptual framework of effective field theories (EFTs), which is intimately 
tied with the notion of renormalization, was widely regarded as uniquely ingrained in the 
domain of quantum field theories (QFTs). In order to dispel the latter fallacy, however, 
suffice it to acknowledge that `renormalization' can in fact be taken as synonymous with 
`resolution of physical scales'. Once this realization is made, it becomes clear that the 
framework of EFTs is evidently a universal one, and that it can be applied on any wide range of 
physical scales. Moreover, the EFT framework is a solid and powerful one, providing an inherent 
setup strategy and a robust toolbox, geared towards yielding high precision predictions to a 
desirable accuracy.

Indeed, just over a decade ago, an EFT approach to handle gravitational waves (GWs), emitted 
from inspiralling compact binaries, which are analytically treated with post-Newtonian (PN) 
gravity \cite{Blanchet:2013haa}, was put forward by Goldberger et 
al.\cite{Goldberger:2004jt,Goldberger:2007hy,Goldberger:2009qd}. This novel approach has 
resulted, in turn, further applications of EFTs in classical gravity. Notably, significant 
progress was demonstrated in the application of EFTs to higher dimensional gravity in the 
context of large extra 
dimensions\cite{Chu:2006ce,Kol:2007rx,Emparan:2009cs,Gilmore:2009ea,Emparan:2009at}. Further, 
also in the context of GWs, an investigation of an EFT treatment for the extreme mass ratio 
inspiral (EMRI) case of compact binaries, was initiated in Ref.~\refcite{Galley:2008ih}. 
Moreover, within a decade the initial EFT approach to PN gravity has been able -- in 
important sectors of the theory -- to catch up and go beyond the spectacular state of the 
art\cite{Levi:2018nxp}, already accomplished within the traditional framework of General 
Relativity (GR).

In what follows we highlight the main meaningful unique advances made in the EFT approach to PN 
gravity, yet more importantly, we point to the main prospects on a broader scope of using field 
theory to study gravity, stretching from delivering highly demanding accurate predictions for 
real-world GW data, to confronting our fundamental grasp of QFTs and gravity theories at all 
scales.

\section{From Gravitational Waves To Gravity At All Scales}

Significant progress was made via the EFT approach to PN gravity, which is concerned with the 
analytical prediction of the inspiral portion of the GW signal. While this progress is evident 
and encouraging, it crucially serves to illustrate the great potential, still awaiting, in the 
use of field theory to study gravity, with many implications at various levels. First, there 
are clear anticipated developments still within the EFTs of PN gravity and the theoretical 
modeling of the GW signal. Further, even if still only in the context of GWs, modern advances 
in scattering amplitudes hold the promise to possibly enable an analytical treatment of the GW 
signal in its entirety, that is also in the strong field regime, beyond the inspiral phase. 
This may significantly improve upon the current phenomenological GW modeling, which relies 
heavily on the effective-one-body (EOB) formulation \cite{Buonanno:1998gg}, by allowing for a 
smooth analytical model of the whole signal. Yet, more broadly, these advances from scattering 
amplitudes also importantly allow us to study gauge and gravity theories at a fundamental 
level, across the classical and quantum regimes. 

Let us first review the main unique advancements accomplished within the EFTs of PN gravity so 
far. The most notable and extensive progress was realized in the treatment of the spinning 
sector, with several new higher order PN corrections. Next-to-leading order (NLO) effects in 
the conservative sector were first tackled in 
Refs.~\refcite{Porto:2006bt,Porto:2008tb,Porto:2008jj}, following 
Refs.~\refcite{Porto:2005ac,Porto:2008tb}, and further resolved and extended to the current 
state of the art, at the fourth PN (4PN) order for rapidly rotating compact objects, in 
Refs.~\refcite{Levi:2008nh,Levi:2015msa,Levi:2011eq,Levi:2014sba,Levi:2014gsa,Levi:2015ixa,Levi:2016ofk}, 
following Refs.~\refcite{Levi:2008nh,Levi:2010zu,Levi:2014sba,Levi:2015msa}. Further, partial 
results at NLO in the radiative sector were presented in Ref.~\refcite{Porto:2010zg}. Moreover, 
following Ref.~\refcite{Galley:2009px}, which also led to a new finite size correction of 
radiation reaction in classical electrodynamics\cite{Galley:2010es}, and  
Refs.~\refcite{Galley:2012hx,Galley:2014wla}, a leading order effect of radiation reaction with 
spins was obtained in Ref.~\refcite{Maia:2017yok}. All in all, beyond the specific new PN 
results obtained with spins, there has been an improvement in the understanding of classical 
spins in gravity. 

Another remarkable finding in the field was the uncovering of classical renormalization group 
(RG) flows of the Wilson coefficients, which characterize the effective theories, both at the  
one-particle level\cite{Kol:2011vg}, and at the level of the composite system, for the 
multipole moments of the binary\cite{Goldberger:2009qd,Goldberger:2012kf}. These give rise to 
higher order PN logarithm corrections, which constitute unique predictions of the EFT 
framework. In addition, the 2PN order correction for a generic $n$-body problem was considered 
in Ref.~\refcite{Chu:2008xm}, where automated computations were first advocated. Finally, the 
`EFTofPNG' code for high precision Feynman calculations in PN gravity was 
created\cite{Levi:2017kzq}, comprising the first comprehensive code in PN theory made public. 
The code consists of independent modules for easy adaptation and future  development, where 
some prospects of building on this state of the art PN technology for various research 
purposes, were outlined in Ref.~\refcite{Levi:2018stw}.

Within the methodology of the EFTs of PN gravity, there are several important directions, where 
development is required. The first is the solution of higher order PN equations of motion 
(EOMs) for accurate orbital dynamics. One interesting related way to find such closed form 
solutions, is the dynamical RG method, suggested in Ref.~\refcite{Galley:2016zee}. Further, the 
treatment of non-conservative sectors with the EFTs of PN gravity has been rather limited, and 
so these various sectors should be tackled with a proper EFT formulation, and implementation 
for new PN results. Another objective to follow is the possible improvement or extension of the 
EFT formulation for spinning objects. For example, an alternative EFT formulation, which is 
ideally suited for the treatment of slowly rotating objects, based on a coset construction, was 
presented in Ref.~\refcite{Delacretaz:2014oxa}. Further investigation is required in order to 
see whether these independent EFT formulations may possibly be integrated, and entail an even 
better understanding of classical spins in gravity. Next, we note the analysis of the mass- and 
spin-induced Wilson coefficients, and of the binary multipole coefficients, which also mostly 
remains to be approached via formal EFT matching. Finally, the continual public development of 
the `EFTofPNG' code, or similar open codes, is required in order to keep up to date with the 
actual progress in PN theory. In that regard, it is important to note that classical theories 
of gravity, which modify GR in the IR, motivated by the cosmological constant problem and the 
dark matter puzzle\cite{Clifton:2011jh,Deffayet:2013lga,deRham:2014zqa,Joyce:2014kja}, can be 
incorporated in a rather straightforward manner into the `EFTofPNG' code, in order to study 
their analytical predictions for the GW signal from the compact binary inspiral, see, e.g., 
Refs.~\refcite{Chu:2012kz,deRham:2012fw,deRham:2012fg}. 

\subsection{Advances and prospects for gravity in field theory}

Ultimately, we would like to invoke the correspondence -- at any level -- between gauge and 
gravity theories, in order to push beyond the state of the art in high precision computation 
for concrete analytical predictions of the real-world GW signal, as well as to directly 
ameliorate our fundamental understanding of the foundations of these theories, in particular of 
gravity. These objectives can be considerably facilitated by turning to modern field theory 
advances in the domain of scattering amplitudes, see, e.g., Ref.~\refcite{Elvang:2015rqa}. In 
fact, the field of EFTs in PN gravity, on which we have focused hitherto, and of scattering 
amplitudes, which are both directly concerned with baldly tackling demanding high precision 
computation driven by experiment, share more profound parallels beyond the obvious robust 
technical tools they entail, see, e.g., in Ref.~\refcite{Smirnov:2012gma}. Both of these fields 
push to the exposure of the universal commonalities across classical and quantum field 
theories, and drive us to confront our fundamental grasp of the underlying foundations of these 
theories. One crucial difference to note, though, between these fields, is that where in 
scattering amplitudes the computational outcome is directly related with the physical 
observables, the objects which are commonly computed in classical gravity are coordinate/gauge 
dependent quantities.

Let us then note, more specifically, some prospective avenues to deploy this broad 
correspondence of gauge theories and gravity. First, there is the standard working knowledge, 
which can be exchanged between the theories, e.g.~multi-loop techniques in QFT, such as 
integration by parts (IBP), and high loop master integrals, along with other Feynman calculus 
and technology\cite{Smirnov:2012gma,Levi:2017kzq,Levi:2018stw}. This was nicely demonstrated, 
e.g., in Refs.~\refcite{Foffa:2016rgu,Damour:2017ced}, where an analytic evaluation of a 
four-loop master integral was provided for the first time from the classical gravity context. 
Furthermore, modern scattering amplitudes advances, such as the BCFW on-shell recursion 
relations\cite{Britto:2004ap}, and generalized unitarity methods, which imply that tree level 
data encodes all multiplicity at the integrand level, were put forward to extract classical 
higher order loop results for gravitational 
scattering\cite{Bjerrum-Bohr:2013bxa,Cachazo:2017jef,Guevara:2017csg,Bjerrum-Bohr:2018xdl}. 
Such a scattering treatment may also enable to analytically tackle the strong field regime of 
the GW signal, and hence smoothly model analytically the entire signal, as we noted above. 

Yet, of particular interest are the novel intriguing color-kinematics or BCJ duality relations, 
and the related double copy correspondence \cite{Bern:2008qj,Bern:2010ue}, see, e.g., review in 
Ref.~\refcite{Carrasco:2015iwa}, which were discovered in the context of high loop computations 
of amplitudes in supersymmetry and supergravity theories. Such relations were already known 
from string theory to hold at tree level, but formulated in terms of the novel generic double 
copy correspondence, these relations have been successfully used to study UV divergences of 
supergravity theories. This recent new perspective on gravity, viewing gravitons as double 
copies of gluons, suggests that even the simplest gauge theories and GR are in fact frameworks, 
which are intimately connected. Though this correspondence was uncovered in the perturbative 
context, it was also found, remarkably, that particular classical \textit{exact} solutions in 
GR, are in fact double copies of exact ``single copy'' counterparts in corresponding gauge 
theories. Such classical solutions include all vacuum stationary solutions, most notably, Kerr 
black holes, as well as their higher dimensional generalizations, see 
Ref.~\refcite{Luna:2016due} and references therein. Yet, the underlying origin of these 
perturbative relations, as well as their essential connection to the particular classical exact 
correspondence revealed, are yet to be uncovered. 

In conjunction with the recent novel methods from scattering amplitudes, it is expected that 
this double copy correspondence can be used to advance the analytical calculations for the 
theoretical prediction of the GW signal, which may also help to shed more light on the nature 
and origin of this correspondence. Related with that end, an important time dependent case was 
studied in terms of the exact double copy of the Kerr-Schild form, of an arbitrarily 
accelerating point source \cite{Luna:2016due}. In this case the radiation current is double 
copied to the radiation stress-energy tensor in Fourier space, and both are related to the 
corresponding scattering amplitudes, which can be obtained from the amputated currents. Hence, 
the work in Ref.~\refcite{Luna:2016due} made a first explicit connection between the classical 
double copy in an exact form, to that in the perturbative scattering amplitudes context. At 
this point it should also be stressed, that from the scattering amplitudes context, it is 
already known that gluon amplitudes can double copy to arbitrary combinations of amplitudes for 
gravitons, and the additional unobserved dilaton and B fields, depending on the choice of the 
polarization states in the gauge theory amplitudes. This was indeed an ambiguous issue in 
Ref.~\refcite{Goldberger:2016iau}, which built on Ref.~\refcite{Luna:2016due}, and that 
subsequently Ref.~\refcite{Luna:2017dtq} set out to address, with Ref.~\refcite{Shen:2018ebu} 
following up successfully at NLO. 

Finally, we note the topic of soft graviton theorems for scattering amplitudes, which were 
recently demonstrated to be equivalent to gravitational memory effects, as part of a triple 
equivalence of the IR structure of gauge and gravity theories among soft theorems, asymptotic 
symmetry, and memory effects\cite{Strominger:2017zoo}. The latter may have observable 
signatures on the GW signal.

\section*{Acknowledgments}

I am grateful to John Joseph Carrasco for his meaningful encouragement.
I would like to thank Donato Bini for the warm hospitality throughout the MG15 meeting, and in 
particular on the session, where this review was presented.
It is also a delightful pleasure to acknowledge Roy Kerr who graced us 
with his inspiring presence.
My work is supported by the European Research Council under the European Union's Horizon 2020 
Framework Programme FP8/2014-2020 ``preQFT'' grant no.~639729, ``Strategic Predictions for 
Quantum Field Theories'' project.


\begin{thebibliography}{69} 

\bibitem{Blanchet:2013haa}
L.~Blanchet, 
{\it {Gravitational Radiation from Post-Newtonian Sources and Inspiralling Compact Binaries}}, 
{\em Living Rev.Rel.} {\bf 17} (2014).

\bibitem{Goldberger:2004jt}
W.~D. Goldberger and I.~Z. Rothstein, 
{\it {An Effective field theory of gravity for extended objects}},  
{\em Phys.Rev.} {\bf D73} (2006) 104029.

\bibitem{Goldberger:2007hy}
W.~D. Goldberger, 
{\it {Les Houches lectures on effective field theories and gravitational radiation}},  
in {\em {Les Houches Summer School - Session 86:
Particle Physics and Cosmology: The Fabric of Spacetime Les Houches, France,
July 31-August 25, 2006}}, 2007.

\bibitem{Goldberger:2009qd}
W.~D. Goldberger and A.~Ross, 
{\it {Gravitational radiative corrections from effective field theory}},  
{\em Phys.Rev.} {\bf D81} (2010) 124015.

\bibitem{Chu:2006ce}
Y.-Z. Chu, W.~D. Goldberger, and I.~Z. Rothstein, 
{\it {Asymptotics of d-dimensional Kaluza-Klein black holes: Beyond the Newtonian 
approximation}},
{\em JHEP} {\bf 03} (2006) 013.

\bibitem{Kol:2007rx}
B.~Kol and M.~Smolkin, 
{\it {Classical Effective Field Theory and Caged Black Holes}},  
{\em Phys. Rev.} {\bf D77} (2008) 064033.

\bibitem{Emparan:2009cs}
R.~Emparan, T.~Harmark, V.~Niarchos, and N.~A. Obers, 
{\it {World-Volume Effective Theory for Higher-Dimensional Black Holes}},  
{\em Phys. Rev. Lett.} {\bf 102} (2009) 191301.

\bibitem{Gilmore:2009ea}
J.~B. Gilmore, A.~Ross, and M.~Smolkin, 
{\it {Caged black hole thermodynamics: Charge, the extremal limit, and finite size effects}},  
{\em JHEP} {\bf 09} (2009) 104. 

\bibitem{Emparan:2009at}
R.~Emparan, T.~Harmark, V.~Niarchos, and N.~A. Obers, 
{\it {Essentials of Blackfold Dynamics}},  
{\em JHEP} {\bf 03} (2010) 063.

\bibitem{Galley:2008ih}
C.~R. Galley and B.~Hu, 
{\it {Self-force on extreme mass ratio inspirals via curved spacetime effective field 
theory}},  
{\em Phys.Rev.} {\bf D79} (2009) 064002. 

\bibitem{Levi:2018nxp}
M.~Levi, {\tt arXiv:1807.01699} 
{\it{Effective Field Theories of Post-Newtonian Gravity: A comprehensive review}}, 2018.

\bibitem{Buonanno:1998gg}
A.~Buonanno and T.~Damour, 
{\it {Effective one-body approach to general relativistic two-body dynamics}},  
{\em Phys.Rev.} {\bf D59} (1999) 084006.

\bibitem{Porto:2006bt}
R.~A. Porto and I.~Z. Rothstein, 
{\it {The Hyperfine Einstein-Infeld-Hoffmann potential}},  
{\em Phys.Rev.Lett.} {\bf 97} (2006) 021101.

\bibitem{Porto:2008tb}
R.~A. Porto and I.~Z. Rothstein, 
{\it {Spin(1)Spin(2) Effects in the Motion of Inspiralling Compact Binaries at Third Order in 
the Post-Newtonian Expansion}},  
{\em Phys.Rev.} {\bf D78} (2008) 044012,
[\textit{Erratum-ibid.\,} D{\bf 81} (2010) 029904].

\bibitem{Porto:2008jj}
R.~A. Porto and I.~Z. Rothstein, 
{\it {Next to Leading Order Spin(1)Spin(1) Effects in the Motion of Inspiralling Compact 
Binaries}},  
{\em Phys.Rev.} {\bf D78} (2008) 044013,
[\textit{Erratum-ibid.\,} D{\bf 81} (2010) 029905].

\bibitem{Porto:2005ac}
R.~A. Porto, 
{\it {Post-Newtonian corrections to the motion of spinning bodies in NRGR}},  
{\em Phys.Rev.} {\bf D73} (2006) 104031.

\bibitem{Levi:2008nh}
M.~Levi, 
{\it {Next to Leading Order gravitational Spin1-Spin2 coupling with Kaluza-Klein reduction}},
{\em Phys.Rev.} {\bf D82} (2010) 064029.

\bibitem{Levi:2015msa}
M.~Levi and J.~Steinhoff, 
{\it {Spinning gravitating objects in the effective field theory in the post-Newtonian 
		scheme}},  
{\em JHEP} {\bf 09} (2015) 219.

\bibitem{Levi:2011eq}
M.~Levi, 
{\it {Binary dynamics from spin1-spin2 coupling at fourth post-Newtonian order}},  
{\em Phys.Rev.} {\bf D85} (2012) 064043.

\bibitem{Levi:2014sba}
M.~Levi and J.~Steinhoff, 
{\it {Equivalence of ADM Hamiltonian and Effective Field Theory approaches at 
next-to-next-to-leading order spin1-spin2 coupling of binary inspirals}},  
{\em JCAP} {\bf 1412} (2014) 003.

\bibitem{Levi:2014gsa}
M.~Levi and J.~Steinhoff, 
{\it {Leading order finite size effects with spins for inspiralling compact binaries}},  
{\em JHEP} {\bf 06} (2015) 059.

\bibitem{Levi:2015ixa}
M.~Levi and J.~Steinhoff, 
{\it {Next-to-next-to-leading order gravitational spin-squared potential via the effective 
field theory for spinning objects in the post-Newtonian scheme}},  
{\em JCAP} {\bf 1601} (2016) 008.

\bibitem{Levi:2016ofk}
M.~Levi and J.~Steinhoff, {\tt arXiv:1607.04252} 
{\it {Complete conservative dynamics for inspiralling compact binaries with spins at fourth 
post-Newtonian order}}, 2016.

\bibitem{Levi:2010zu}
M.~Levi, 
{\it {Next to Leading Order gravitational Spin-Orbit coupling in an Effective Field Theory 
approach}},  
{\em Phys.Rev.} {\bf D82} (2010) 104004.

\bibitem{Porto:2010zg}
R.~A. Porto, A.~Ross, and I.~Z. Rothstein, 
{\it {Spin induced multipole moments for the gravitational wave flux from binary inspirals to 
third Post-Newtonian order}},  
{\em JCAP} {\bf 1103} (2011) 009.

\bibitem{Galley:2009px}
C.~R. Galley and M.~Tiglio, 
{\it {Radiation reaction and gravitational waves in the effective field theory approach}},  
{\em Phys.Rev.} {\bf D79} (2009) 124027. 

\bibitem{Galley:2010es}
C.~R. Galley, A.~K. Leibovich, and I.~Z. Rothstein, 
{\it {Finite size corrections to the radiation reaction force in classical electrodynamics}},
{\em Phys.Rev.Lett.} {\bf 105} (2010) 094802.

\bibitem{Galley:2012hx}
C.~R. Galley, 
{\it {Classical Mechanics of Nonconservative Systems}},  
{\em Phys.Rev.Lett.} {\bf 110} (2013) 174301.

\bibitem{Galley:2014wla}
C.~R. Galley, D.~Tsang, and L.~C. Stein, {\tt arXiv:1412.3082}
{\it {The principle of stationary nonconservative action for classical mechanics and field 
theories}}, 2014.

\bibitem{Maia:2017yok}
N.~T. Maia, C.~R. Galley, A.~K. Leibovich, and R.~A. Porto, 
{\it {Radiation reaction for spinning bodies in effective field theory II: Spin-spin 
effects}},  
{\em Phys. Rev.} {\bf D96} (2017) 084065.

\bibitem{Kol:2011vg}
B.~Kol and M.~Smolkin, 
{\it {Black hole stereotyping: Induced gravito-static polarization}},  
{\em JHEP} {\bf 02} (2012) 010.

\bibitem{Goldberger:2012kf}
W.~D. Goldberger, A.~Ross, and I.~Z. Rothstein, 
{\it {Black hole mass dynamics and renormalization group evolution}},  
{\em Phys. Rev.} {\bf D89} (2014) 124033.

\bibitem{Chu:2008xm}
Y.-Z. Chu, 
{\it {The n-body problem in General Relativity up to the second post-Newtonian order from 
perturbative field theory}},  
{\em Phys.Rev.} {\bf D79} (2009) 044031.

\bibitem{Levi:2017kzq}
M.~Levi and J.~Steinhoff, 
{\it {EFTofPNG: A package for high precision computation with the Effective Field Theory of 
Post-Newtonian Gravity}},
{\em Class. Quant. Grav.} {\bf 34} (2017) 244001.

\bibitem{Levi:2018stw}
M.~Levi, {\tt arXiv:1811.12401} 
{\it{A public framework for Feynman calculations and post-Newtonian gravity}}, 2018.


\bibitem{Galley:2016zee}
C.~R. Galley and I.~Z. Rothstein, 
{\it {Deriving analytic solutions for compact binary inspirals without recourse to adiabatic 
approximations}},  
{\em Phys. Rev.} {\bf D95} (2017) 104054.

\bibitem{Delacretaz:2014oxa}
L.~V. Delacr\'etaz, S.~Endlich, A.~Monin, R.~Penco, and F.~Riva, 
{\it {(Re-)Inventing the Relativistic Wheel: Gravity, Cosets, and Spinning Objects}},  
{\em JHEP} {\bf 11} (2014) 008.

\bibitem{Clifton:2011jh}
T.~Clifton, P.~G. Ferreira, A.~Padilla, and C.~Skordis, 
{\it {Modified Gravity and Cosmology}},  
{\em Phys. Rept.} {\bf 513} (2012) 1--189.

\bibitem{Deffayet:2013lga}
C.~Deffayet and D.~A. Steer, 
{\it {A formal introduction to Horndeski and Galileon theories and their generalizations}}, 
{\em Class. Quant. Grav.} {\bf 30} (2013) 214006.

\bibitem{deRham:2014zqa}
C.~de~Rham, 
{\it {Massive Gravity}}, 
{\em Living Rev. Rel.} {\bf 17} (2014) 7.

\bibitem{Joyce:2014kja}
A.~Joyce, B.~Jain, J.~Khoury, and M.~Trodden, 
{\it {Beyond the Cosmological Standard Model}},  
{\em Phys. Rept.} {\bf 568} (2015) 1--98.

\bibitem{Chu:2012kz}
Y.-Z. Chu and M.~Trodden, 
{\it {Retarded Green’s function of a Vainshtein system and Galileon waves}},  
{\em Phys. Rev.} {\bf D87} (2013) 024011.

\bibitem{deRham:2012fw}
C.~de~Rham, A.~J. Tolley, and D.~H. Wesley, 
{\it {Vainshtein Mechanism in Binary Pulsars}},  
{\em Phys. Rev.} {\bf D87} (2013) 044025.

\bibitem{deRham:2012fg}
C.~de~Rham, A.~Matas, and A.~J. Tolley, 
{\it {Galileon Radiation from Binary Systems}},  
{\em Phys. Rev.} {\bf D87} (2013) 064024.

\bibitem{Elvang:2015rqa} 
H.~Elvang and Y.~t.~Huang, 
{\em {Scattering Amplitudes in Gauge Theory and Gravity}}
\newblock (Cambridge University Press, 2015).

\bibitem{Smirnov:2012gma} 
V.~A.~Smirnov, 
{\em {Analytic tools for Feynman integrals,}}
\newblock Springer Tracts Mod.\ Phys.\  {\bf 250}, 1 (2012).

\bibitem{Foffa:2016rgu}
S.~Foffa, P.~Mastrolia, R.~Sturani, and C.~Sturm, 
{\it {Effective field theory approach to the gravitational two-body dynamics, at fourth 
post-Newtonian order and quintic in the Newton constant}},  
{\em Phys. Rev.} {\bf D95} (2017) 104009. 

\bibitem{Damour:2017ced}
T.~Damour and P.~Jaranowski, 
{\it {Four-loop static contribution to the gravitational interaction potential of two point 
masses}},  
{\em Phys. Rev.} {\bf D95} (2017) 084005. 

\bibitem{Britto:2004ap}
R.~Britto, F.~Cachazo, and B.~Feng, 
{\it {New recursion relations for tree amplitudes of gluons}},  
{\em Nucl. Phys.} {\bf B715} (2005) 499--522.

\bibitem{Bjerrum-Bohr:2013bxa}
N.~E.~J. Bjerrum-Bohr, J.~F. Donoghue, and P.~Vanhove, 
{\it {On-shell Techniques and Universal Results in Quantum Gravity}},  
{\em JHEP} {\bf 02} (2014) 111. 

\bibitem{Cachazo:2017jef}
F.~Cachazo and A.~Guevara, {\tt arXiv:1705.10262}
{\it {Leading Singularities and Classical Gravitational Scattering}}, 2017.

\bibitem{Guevara:2017csg}
A.~Guevara, {\tt arXiv:1706.02314}
{\it {Holomorphic Classical Limit for Spin Effects in Gravitational and Electromagnetic 
Scattering}}, 2017.

\bibitem{Bjerrum-Bohr:2018xdl} 
N.~E.~J. Bjerrum-Bohr, P.~H. Damgaard, G.~Festuccia, L.~Plant\'e, and P.~Vanhove, 
{\it {General Relativity from Scattering Amplitudes}}, 
{\em Phys. Rev. Lett.} {\bf 121} (2018) 171601.

\bibitem{Bern:2008qj}
Z.~Bern, J.~J.~M. Carrasco, and H.~Johansson, 
{\it {New Relations for Gauge-Theory Amplitudes}},  
{\em Phys. Rev.} {\bf D78} (2008) 085011.

\bibitem{Bern:2010ue}
Z.~Bern, J.~J.~M. Carrasco, and H.~Johansson, 
{\it {Perturbative Quantum Gravity as a Double Copy of Gauge Theory}},  
{\em Phys. Rev. Lett.} {\bf 105} (2010) 061602.

\bibitem{Carrasco:2015iwa}
J.~J.~M. Carrasco, 
{\it {Gauge and Gravity Amplitude Relations}},  
in {\em{Proceedings, Theoretical Advanced Study Institute in Elementary Particle Physics: 
Journeys Through the Precision Frontier: Amplitudes for Colliders
(TASI 2014): Boulder, Colorado, June 2-27, 2014}}, pp.~477--557, WSP, 2015.

\bibitem{Luna:2016due}
A.~Luna, R.~Monteiro, I.~Nicholson, D.~O'Connell, and C.~D. White, 
{\it {The double copy: Bremsstrahlung and accelerating black holes}},  
{\em JHEP} {\bf 06} (2016) 023. 

\bibitem{Goldberger:2016iau}
W.~D. Goldberger and A.~K. Ridgway, 
{\it {Radiation and the classical double copy for color charges}},  
{\em Phys. Rev.} {\bf D95} (2017) 125010.

\bibitem{Luna:2017dtq}
A.~Luna, I.~Nicholson, D.~O'Connell, and C.~D. White, 
{\it {Inelastic Black Hole Scattering from Charged Scalar Amplitudes}},  
{\em JHEP} {\bf 03} (2018) 044. 

\bibitem{Shen:2018ebu}
C.-H. Shen, 
{\it {Gravitational Radiation from Color-Kinematics Duality}}, 
{\em JHEP} {\bf 11} (2018) 162. 

\bibitem{Strominger:2017zoo}
A.~Strominger, {\tt arXiv:1703.05448}
{\it {Lectures on the Infrared Structure of Gravity and Gauge Theory}}, 2017. 

\end{thebibliography}
\end{document}